\documentclass[preprint,12pt]{elsarticle}
\usepackage{graphicx}
\usepackage{amssymb, amsmath}
\usepackage{lineno}
\usepackage{upgreek}
\usepackage{color}
\usepackage{hyperref}

\newcommand{\Nuc}[2]{\ensuremath{^{#2}\mbox{#1}}}
\newcommand{\scnota}[2]{${#1}\times 10^{#2}$}

\begin{document}
\begin{frontmatter}

\title{A search for rare and induced nuclear decays in hafnium}

\author[queens]{B.~Broerman\corref{cor1}}
\cortext[cor1]{broerman@owl.phy.queensu.ca}
\author[lngs]{M.~Laubenstein}
\author[queens]{S.~Nagorny}
\author[queens,mi,perimeter]{N.~Song}
\author[queens,mi,perimeter]{A.C.~Vincent}

\address[queens]{Department of Physics, Enigneering Physics and Astronomy, Queen's University, Kingston, ON, K7L 3N6, Canada}
\address[lngs]{INFN -- Laboratori Nazionali del Gran Sasso, Assergi, I-67100, Italy}
\address[mi]{Arthur B. McDonald Canadian Astroparticle Physics Research Institute, Kingston ON K7L 3N6, Canada}
\address[perimeter]{Perimeter Institute for Theoretical Physics, Waterloo ON N2L 2Y5, Canada}

\begin{abstract}
A measurement of hafnium foil using a modified ultra-low-background high purity detector with optimized sample-to-detector geometry was performed at Laboratori Nazionale del Gran Sasso. Radiopurity of the stock Hf foil was studied in detail, in addition to an analysis of data collected over 310~days to search for rare processes that can occur in natural Hf isotopes. Firstly, limits on alpha decays of all natural Hf isotopes to the first excited state of the daughter nuclides were established in the range of $10^{16}$--$10^{18}$a (90\% C.L.). Secondly, a search for modes of double electron capture and electron capture with positron emission in \Nuc{Hf}{174} was performed, yielding half-life limits $10^{16}$--$10^{18}$a (90\% C.L.). Lastly, novel dark matter-induced nuclear excitations in hafnium isotopes were investigated. For dark matter with 1~TeV/$c^2$ mass, leading limits on the inelastic dark matter--nucleon cross section are set for mass splittings in the range 428~keV $< \delta <$ 473~keV.  
\end{abstract}

\begin{keyword}
rare alpha decay \sep double beta decay \sep dark matter \sep low-background measurements \sep Hf isotopes \sep $^{174}$Hf isotope
\end{keyword}
\end{frontmatter}

\section{Introduction}
The radioactive decays of many natural isotopes, despite being energetically allowed, are not easily observed when the expected half-life is much greater than the age of the Universe ($\mathcal{O}(10^{10})$~a)~\cite{cayrel2001measurement}. Such decays obey conservation laws and are well-described by conventional nuclear processes without necessarily invoking more exotic physics beyond the Standard Model like neutrinoless double beta decay~\cite{vergados2012theory, giuliani2012neutrinoless, gomez2012search}. 

Searches for these rare nuclear processes have become more practical in part due to the usage of high-purity materials and enriched isotopes, operation of cryogenic (scintillating-\cite{casali2016cryogenic}) bolometers~\cite{casali2014discovery, cozzini2004detection, de2003experimental} and ultra-low-background high purity germanium detectors (ULB HPGe)~\cite{laubenstein2019new}, as well as the requirement that these measurements be performed in underground laboratories screening against cosmic ray backgrounds. Proper application of these techniques have allowed for sensitivity to half-lives reaching $10^{16}-10^{19}$~a.

In many cases, high decay energy processes allow transitions not only directly to the ground states (g.s.) but also via excited levels of the daughter nuclei. This presents a new signature for rare decay searches by measuring the deexcitation photons. If the decay is undergoing a transition to a low-lying excited state (i.e.$ <250$~keV), there are however additional challenges to overcome when attempting to detect these photons with a HPGe detector. As event rates from rare decays with $T_{1/2}\gtrsim10^{14}$~a are of order one per day, experimental approaches must maximize the signal to background rate. Indeed, increasing the detection efficiency and minimizing self-absorption of the emitted low-energy photons within the sample and material surrounding the HPGe detector can be more effective than simply reducing backgrounds by sample purification as the sensitivity to $T_{1/2} \propto \rm{efficiency}\cdot (background)^{-1/2}$; an increase in efficiency can lead to a higher $T_{1/2}$ sensitivity even in the presence of larger background rate. 

Recently, rare $\alpha$ decays of hafnium isotopes to excited states of ytterbium using a highly-purified Hf sample were investigated with two conventional HPGe detectors in coincidence~\cite{danevich2020first}. This measurement was conducted in the HADES underground laboratory. No evidence of such Hf $\alpha$ decays were observed and limits on the corresponding half-lives were placed in the range of $10^{15}-10^{18}$ a. This sensitivity was limited by low detection efficiency in the region of interest due to the detector and sample geometry. 

A modified HPGe detector, as in~\cite{ptpaper}, was used in the experimental search presented here with the copper high voltage contact inside the detector housing replaced by a foil made of the target material. A significant increase in efficiency was achieved by optimizing the thickness of the Hf foil to minimize self-absorption and placing it inside the HPGe detector housing. 

Hafnium is of interest not only for rare $\alpha$ decays, but also has potential in double beta decay searches. One isotope, \Nuc{Hf}{174}, can undergo double electron capture (2$\varepsilon$) and electron capture with positron emission ($\varepsilon\beta^+$)~\cite{wang2017ame2016}. The only experimental search for such decays~\cite{danevich2020firstbeta} was performed by the authors of \cite{danevich2020first} with the same sample and detector technique. Again, no evidence of such $\beta\beta$ decays were observed and limits on the corresponding half-lives were placed in the range of $10^{16}-10^{18}$~a. We summarize the energy released in these rare processes in Table~\ref{tab:elevel}.

\begin{table}[h!]\label{tab:elevel}
\begin{center}
\caption{Decay of the natural isotopes of Hf to the ground state of Yb daughters, natural isotopic abundance $\delta$~\cite{meija2016isotopic}, and Q values~\cite{wang2017ame2016}.}
\begin{tabular}{ r | r | r | r }
Transition&\begin{tabular}{@{}r@{}}Decay Mode \\ (g.s. to g.s.)\end{tabular}&~$\delta$~&~Q [keV]~\\ \hline
\Nuc{Hf}{174}~$\rightarrow$~\Nuc{Yb}{170}	& $\alpha$	& 0.0016(12)	& 2494.5(23) \\
\Nuc{Hf}{174}~$\rightarrow$~\Nuc{Yb}{172} 	& $2\beta$	& 0.0016(12)	& 1100.0(2.3) \\
\Nuc{Hf}{176}~$\rightarrow$~\Nuc{Yb}{172}	& $\alpha$	& 0.0526(70)	& 2254.2(15) \\
\Nuc{Hf}{177}~$\rightarrow$~\Nuc{Yb}{173}	& $\alpha$	& 0.1860(16)	& 2245.7(14) \\
\Nuc{Hf}{178}~$\rightarrow$~\Nuc{Yb}{174}	& $\alpha$	& 0.2728(28)	& 2084.4(14) \\
\Nuc{Hf}{179}~$\rightarrow$~\Nuc{Yb}{175}	& $\alpha$	& 0.1362(11)	& 1807.7(14) \\
\Nuc{Hf}{180}~$\rightarrow$~\Nuc{Yb}{176}	& $\alpha$	& 0.3508(33)	& 1287.1(14) \\
\end{tabular}
\end{center}
\end{table}

Recently, long-lived isomeric states (\Nuc{Ta}{180m}, \Nuc{Hf}{178m2}, etc.) have been proposed as detectors for strongly interacting dark matter (DM) and inelastic dark matter (IDM) particles~\cite{pospelov2019metastable} which, after scattering, initiate the isomeric state decay. Although such a state exists in Hf (\Nuc{Hf}{178m2}), its lifetime is not long enough to perform such an analysis. However, we raise the important point that Hf is heavy enough that collisions with DM can excite certain transitions, which may be recorded via their gamma ray deexictation.  Thus, we are able to probe parts of the DM parameter space that remain inaccessible to conventional direct detection experiments. To our knowledge, this is the first search for dark matter scattering on hafnium.

Results presented here are from a novel search for Hf decays ($\alpha$, $\beta\beta$) to excited states of Yb by measuring with high efficiency the deexcitation gammas and X-rays of Hf isotopes or the daughter nuclei, in addition to a search for DM-induced deexcitations.  We begin in Sec. \ref{sec:setup} with an overview of our experimental setup. Alpha and beta decay searches, are respectively described in Secs. \ref{sec:alpha} and \ref{sec:beta}, and our DM search is detailed in Sec. \ref{sec:dm}. We end with a brief discussion in Sec. \ref{sec:disc} and conclude in Sec. \ref{sec:conc}.

\section{Experimental Setup}\label{sec:setup}
An annealed hafnium foil from Alfa Aesar (stock number 10793, lot H30M31) was used for the measurement with a purity of 99.5\% for metals excluding Zr, which is present at the 1.85\% level. The elemental abundances in parts-per-million (ppm), as provided in the supplier's certificate of analysis~\cite{hfDataSheet}, are shown in Table~\ref{tab:hfCert}. The Hf foil was ultrasonically-cleaned with a neutral soap solution and rinsed with ultra-pure water and alcohol before being mounted in the detector. 

\begin{table}[h!]\label{tab:hfCert}
\begin{center}
\caption{Elemental abundance of the Hf sample~\cite{hfDataSheet}. Abundances are given in ppm unless noted.}
\begin{tabular}{ r r | r r}
Element & Abundance & Element & Abundance \\ \hline
H 	& 	$< 3$ 	& Ni	&	$< 25$ \\
B	&	$< 0.5$	& Cu 	& 	$< 25$ \\			
C	&	55 		& Zr	&	1.85\% \\
N	& 	20		& Nb	&	$< 50$ \\
O	&	$< 230$	& Mo 	& 	$< 10$ \\
Al	&	28		& Cd	& 	$< 2$ \\
Si	&	$< 25$ 	& Sn	&	$< 10$ \\
P	&	$< 3$ 	& Gd 	&	$< 1$ \\   
Ti	&	$< 25$ 	& Ta	&	$< 100$ \\	
V 	& $< 10$ 	& W	&	$< 10$ \\
Cr 	& 	$< 20$ 	& Pb	&	$< 5$ \\
Mn 	& 	$< 20$ 	& Th	&	$< 1$ \\	
Fe 	&	155 	& U		&	$< 1$ \\
Co	&	$< 5$ \\
\end{tabular}
\end{center}
\end{table}

\begin{figure}[h!]
\begin{center}
	\includegraphics[width=0.5\textwidth]{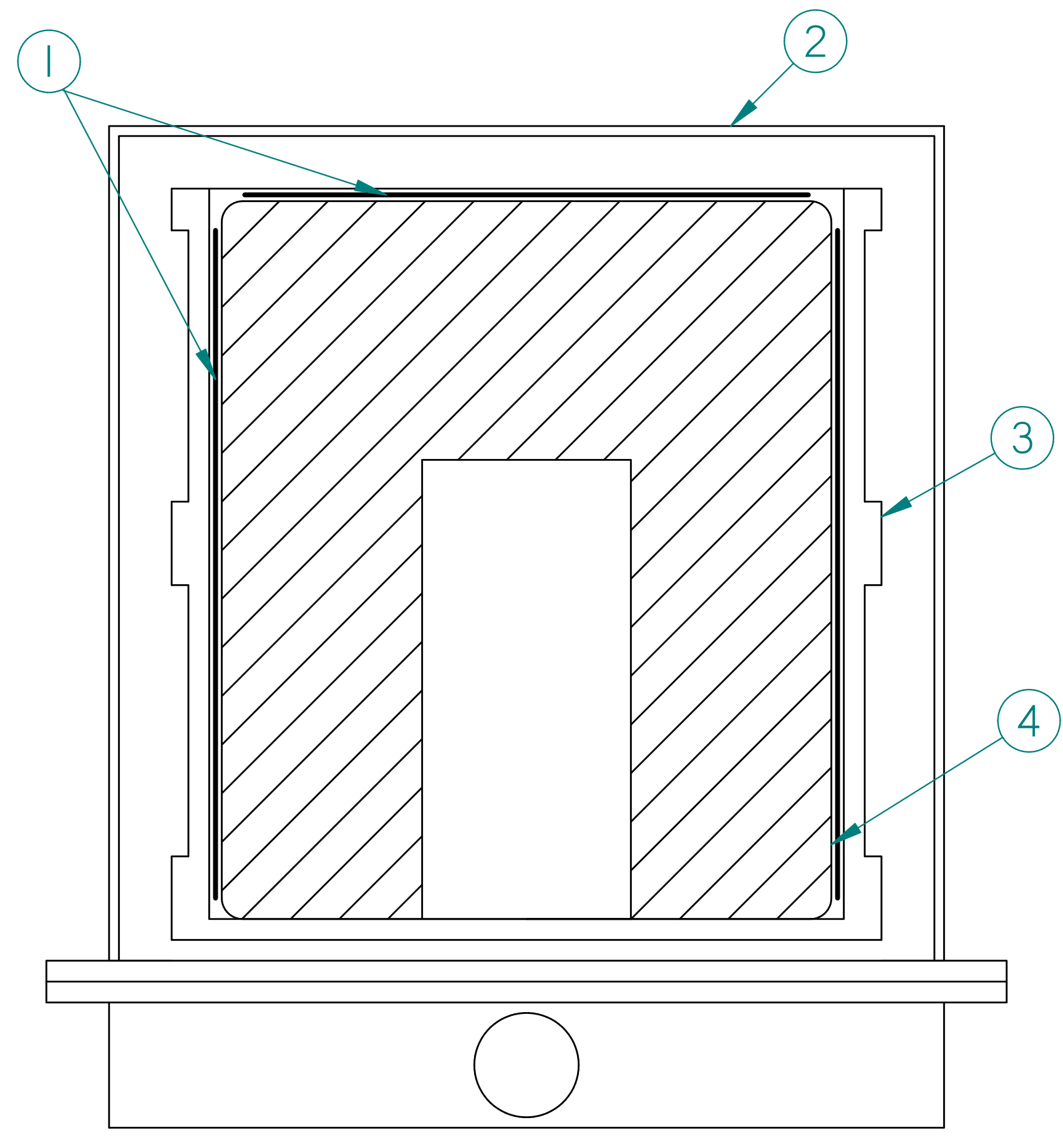}
	\caption{Section view of the detector and sample (not to scale) with 1) hafnium foils on the top and wrapping the Ge crystal acting as the target and high-voltage contact, 2) copper end cap of 1~mm thickness, 3) copper HPGe crystal holder, and 4) HPGe semi-coaxial p-type crystal.}
	\label{fig:foil}
\end{center}

\end{figure}

The data were collected with a HPGe semi-coaxial p-type detector (GS-1) located underground at the Gran Sasso National Laboratory~\cite{ptpaper}. A schematic of the detector and sample is shown in Figure~\ref{fig:foil}. The Ge crystal (4) has a diameter of 70~mm and a height of 70~mm and was surrounded by the hafnium foil (1) with a thickness of $0.25\pm0.01$~mm serving as the high voltage contact and sample. A second disk of Hf foil of the same diameter was placed on top of the crystal to increase coverage and sample mass. The total mass of Hf was $55.379\pm0.001$~g. The foil thickness was selected to minimize self-absorption of low-energy $\gamma$ and X-rays within the sample, optimizing the overall detection efficiency. The crystal and Hf foil were surrounded by an approximately 1-mm-thick sheet of polyethylene for electrical insulation, placed inside the crystal holder (3) made of oxygen-free high conductivity (OFHC) copper, and firmly fixed with screws to guarantee a high quality, stable contact. The end cap (2), made of OFHC copper was then closed and the detector placed in a multi-layer passive shield with overall dimensions $60 \times 60 \times 80$~cm divided into a fixed outer shell and a movable insert. The outer shell is made from low-activity lead with a \Nuc{Pb}{210} activity $< 30$ Bq/kg while the movable insert is made of 5~cm OFHC copper, 7~cm ULB lead (provided by PLOMBUM FL with a \Nuc{Pb}{210} activity $< 6$ Bq/kg), and low-activity lead. The movable insert (inner-most shielding layers and detector) was housed in an acrylic box and continuously flushed with high-purity boil-off nitrogen. 

\begin{figure}[h!]\label{fig:fullSpec}
\begin{center}
	\includegraphics[width=\textwidth]{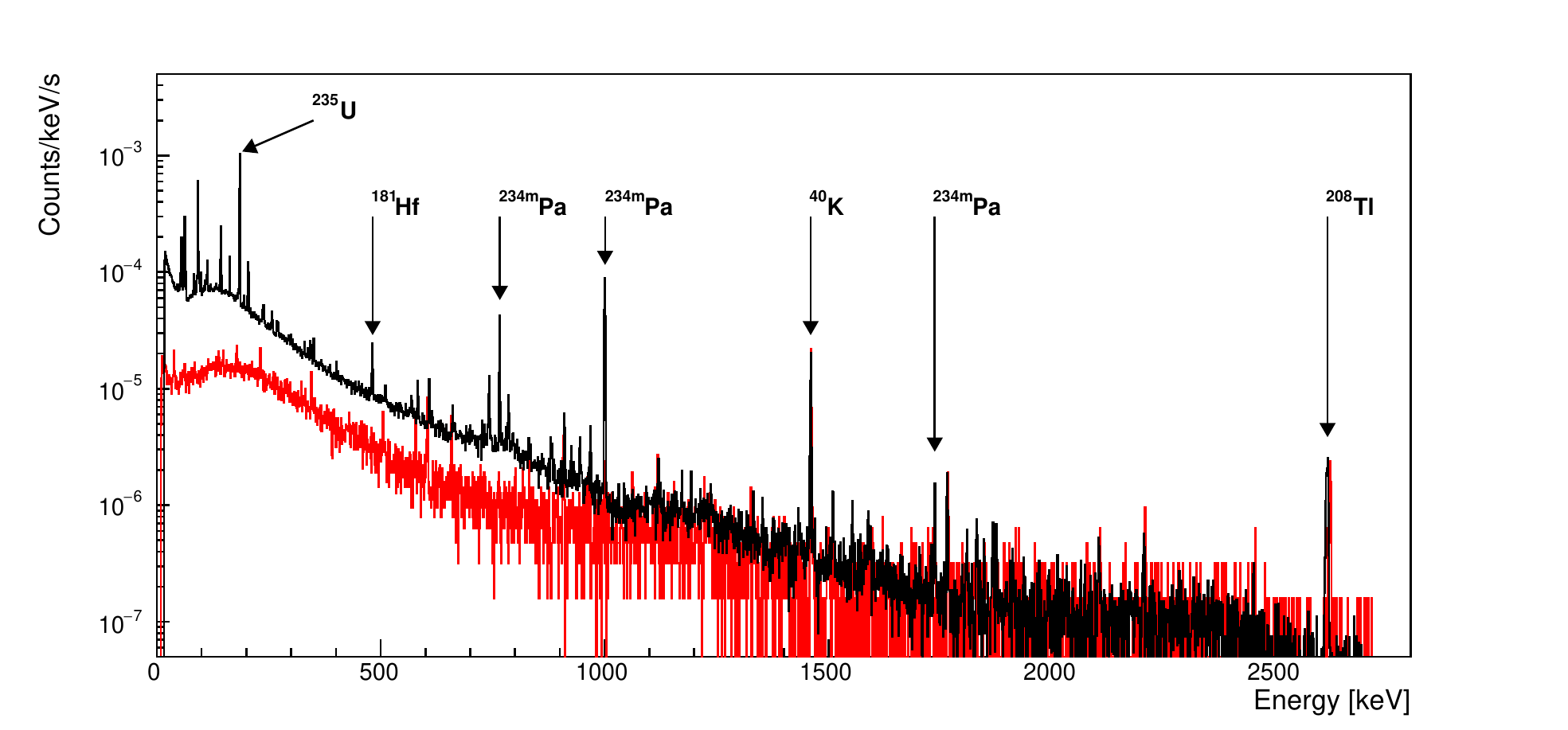}
	\caption{Complete energy spectrum of Hf foil with major lines from contaminant and cosmogenically-activated isotopes indicated (black) and background spectrum using a copper high voltage contact (red). Data was accumulated for approximately 310 days.}
\end{center}
\end{figure}

By using the Hf foil as the high voltage contact for the HPGe detector the sample is placed as close to the detector as possible, increasing the detection efficiency of low-energy $\gamma$ and X-rays. The efficiency at 84~keV is 4.69\% as calculated with a GEANT4 simulation of the detector geometry, with a measured full-width at half-maximum (FWHM) energy resolution of 1.2~keV. The energy resolution at 662~keV was measured to be 1.1~keV FWHM. A total of 0.85102 years of data was accumulated with an 18~keV energy threshold. The full energy spectrum in counts/keV/s is shown in Figure~\ref{fig:fullSpec} with major lines identified, along with a time-normalized background spectrum taken with a copper high voltage contact. 

This value of full energy peak (FEP) efficiency does take into account a precise description of the inhomogeneity of the dead layer in the HPGe crystal. The dead layer was determined through high precision scanning measurements with collimated \Nuc{Am}{241} and \Nuc{Ba}{133} sources all around the crystal.

Radiopurity of the sample was determined by fitting the observed energy spectrum with a gaussian plus linear function around known peaks in the uranium and thorium decay chains (see Figure~\ref{fig:alphaZoom} for a specific example). The activity $\mathcal{A}$ is then calculated $\mathcal{A} = n/(\eta\cdot b\cdot m\cdot t)$ where $n$ is the number of counts above background, $\eta$ is the FEP detection efficiency for a given energy calculated from a GEANT4 simulation, $b$ is the branching ratio for the specific gamma ray emission, $m$ is the mass of the sample and $t$ the data collection time. Activities of prominent isotopes are given in Table~\ref{tab:purity} where the uncertainty is calculated from the statistical uncertainty on signal, background, and detection efficiency. Where no clear peak was identified, 90\% confidence level limits are calculated.

\begin{table}[h!]
\centering
\caption{Measured contaminant activities of the Hf sample. Where no clear peak is observed, limits are placed at 90\% C.L.}
\begin{tabular}{r|r}
Isotope~&~Activity [mBq/kg]~\\ \hline
\Nuc{K}{40}		&	$< 13$	\\
\Nuc{Co}{60}	&	$< 0.78$ \\
\Nuc{Cs}{137}	&	$< 0.70$ \\
\Nuc{Hf}{172} 	& 	$< 7.4$ \\
\Nuc{Hf}{175}	&	7 $\pm$ 1 \\ 
\Nuc{Hf}{178m2}	&   $< 1.7$	\\
\Nuc{Hf}{181}	& 	42 $\pm$ 4	\\
\Nuc{Hf}{182}	& 	$< 3.4$	\\ \hline
\Nuc{Th}{232}:~~\Nuc{Ac}{228}	&	3 $\pm$ 1 \\
				\Nuc{Th}{228}	&	3 $\pm$ 1 \\ \hline
\Nuc{U}{235}:~~~~\Nuc{U}{235}		&	257 $\pm$ 23 \\ 
				\Nuc{Pa}{231}	&	17 $\pm$ 1 \\ 
				\Nuc{Th}{231}   &   200 $\pm$ 100 \\ \hline 
\Nuc{U}{238}:~~\Nuc{Th}{234}    &   4960 $\pm$ 355 \\
                \Nuc{Pa}{234m}	& 	5250 $\pm$ 379 \\
				\Nuc{Bi}{214}	&   $< 2.0$ \\
                \Nuc{Pb}{210}	& 	$<700$ \\
\end{tabular}
\label{tab:purity}
\end{table}

The most prominent lines are from isotopes in the uranium and thorium decay chains. Secular equilibrium is broken early in the decay chains of \Nuc{U}{235} and \Nuc{U}{238}. Notably, a high abundance of \Nuc{Th}{234} (4960 $\pm$ 355 mBq/kg) and \Nuc{Pa}{234m} (5250 $\pm$ 379 mBq/kg) from the top of the \Nuc{U}{238} chain is observed. In addition to decays originating from the uranium and thorium decay chains, cosmogenically-activated \Nuc{Hf}{181} ($T_{1/2} = 42.39$ days) is measured at $42 \pm 4$ mBq/kg. Since the Hf foil is a stock material which has undergone no additional purification it is expected that elevated levels of contamination can be present. 

The ratio of \Nuc{U}{235}/\Nuc{U}{238} is expected to be 0.046 based on natural isotopic abundance. This ratio is measured to be $0.049\pm0.006$ using the activity of \Nuc{Pa}{234m} assuming equilibrium with \Nuc{U}{238}, in good agreement with expectation.

\section{Search for alpha decays to excited levels}\label{sec:alpha}
While all alpha decays of the six natural isotopes of hafnium are investigated in this analysis, the detection efficiency was specifically optimized for the search of the \Nuc{Hf}{174} decay to the 84.3~keV excited state of \Nuc{Yb}{170}. The relevant details of the alpha decays of hafnium to the ground state of the ytterbium daughters are shown Table~\ref{tab:elevel} along with the natural isotopic abundance $\delta$ and Q values, while the low-energy gamma lines in the interval $75-105$~keV used in the search for those Hf isotope decays to the first excited state of Yb are shown in Table~\ref{tab:hfLimits}. 

\begin{figure}[h!]\label{fig:alphaZoom}
\begin{center}
	\includegraphics[width=\textwidth]{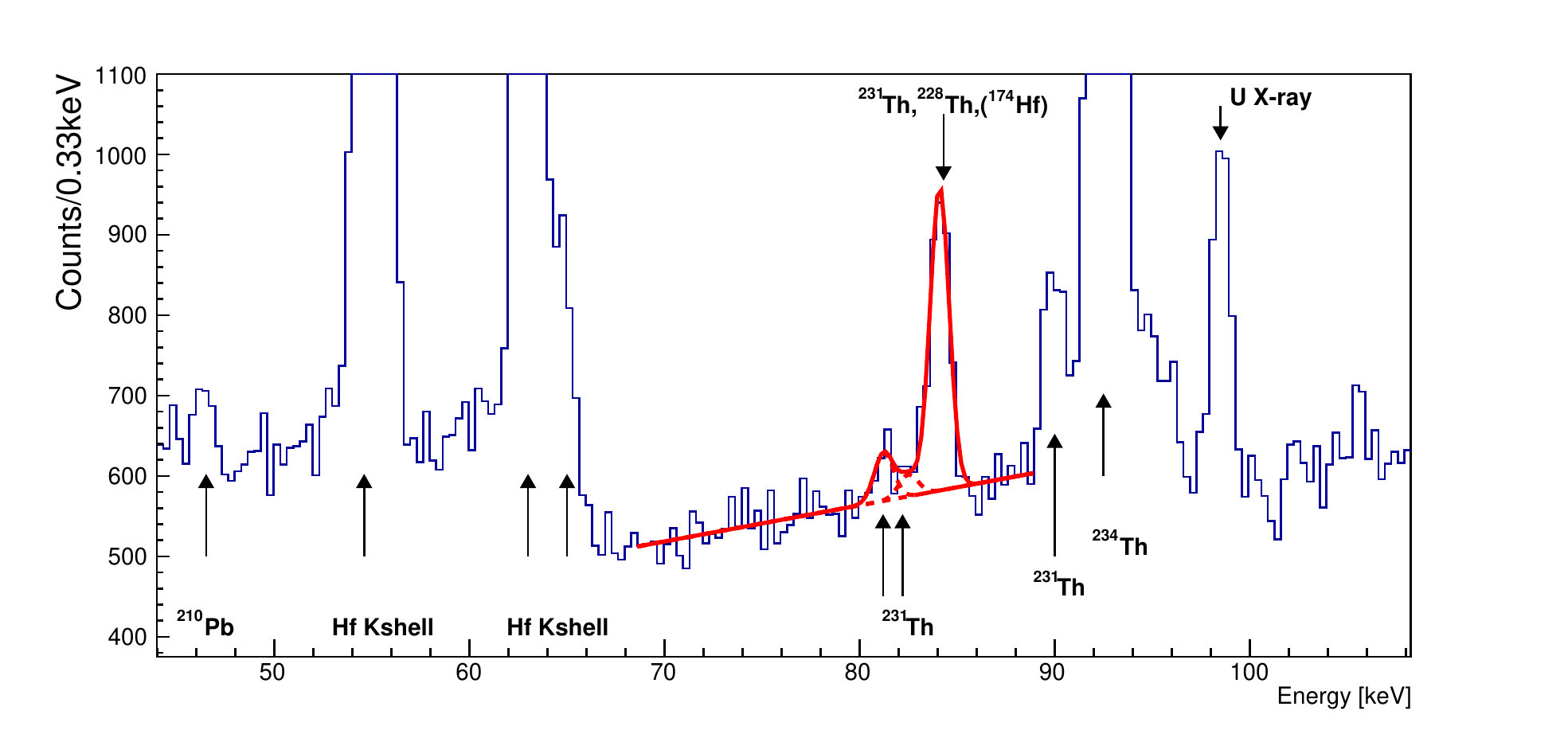}
	\caption{Energy spectrum in the region of interest for Hf $\alpha$ decays to the first excited state of Yb daughter nuclides. A linear + 3 gaussian fit is shown for the 83.4~keV energy expected from the transition \Nuc{Hf}{174}~$\rightarrow$~\Nuc{Yb}{170}.}
\end{center}
\end{figure}

The energy spectrum in the region of interest encompassing these decays ($50-110$~keV) is shown in Figure~\ref{fig:alphaZoom}. A combination of Hf X-ray peaks are observed at 54.6/55.8~keV and 63.0/63.2~keV, along with peaks from the uranium and thorium decay chains. The gamma-ray peaks in the region from $90-100$~keV is due predominantly to \Nuc{Th}{234} in the \Nuc{U}{238} chain around 92~keV and \Nuc{Th}{231} from \Nuc{U}{235} at 93~keV, with additional contributions from \Nuc{Ac}{228}, \Nuc{U}{235}, \Nuc{Th}{227}, and uranium and thorium X-rays.

An example fit to the energy spectrum used in the analysis of the \Nuc{Hf}{174} \Nuc{Hf}{174}~$\rightarrow$ \Nuc{Yb}{170} transition shown in red in Figure~\ref{fig:alphaZoom}. This specific peak is fit with a linear + three gaussian functions ($\chi^2/\rm{n.d.f}$ = 52.4/53) to account for the flat Compton background, \Nuc{Th}{231} lines at 81.2~keV and 82.1~keV, and the main peak backgrounds of \Nuc{Th}{231} (84.2~keV) and \Nuc{Th}{228} (84.4~keV). The widths of the three gaussians were set equal as the energy resolution is not expected to vary over the very limited energy range, and the relative heights of the two background peaks at 81.2~keV and 82.1~keV were fixed based on the ratio of the intensities for each decay. A total of 1493 $\pm$ 27 counts are observed in the main peak at 84~keV, consistent with the expected background of 1540 $\pm$ 40 from \Nuc{Th}{231}, \Nuc{Th}{228} (X-ray \Nuc{Ra}{224}), \Nuc{Ra}{223} (X-ray \Nuc{Rn}{219}) and \Nuc{Th}{234} (X-ray \Nuc{Pa}{234}). We take the number of counts $S$ to be the difference of the observed counts and expected background resulting in $S= -47 \pm 48$ counts. An upper limit on the number of counts, $\lim S$, is then calculated following the Feldman-Cousins method~\cite{feldman1998unified} at 90\% C.L.~to be 40~counts, and the limit on the half-life of this decay of \scnota{2.8}{16}~a is calculated using:

\begin{equation}\label{eq:limitA}
	{\rm{lim}}~T_{1/2} = \ln(2)\cdot \mathcal{N}\cdot y\cdot \eta \cdot t/\lim S,
\end{equation}
where $\mathcal{N}$ is the number of nuclides in the sample, $y$ is the $\gamma$ yield from the first excited state of the daughter nuclide, $\eta$ is the FEP detection efficiency for $E_\gamma$, and $t$ is the counting time. 

Searches for the other $\alpha$ decays were performed with two methods depending on the presence of backgrounds. Conservatively, contributions from the background spectrum are only considered when a prominent gamma line exists in the region of interest. Fits to the spectrum in regions with featureless backgrounds were reduced to linear + single gaussians to describe the flat Compton background and the expected signal from the alpha decay with $\chi^2/$n.d.f.~from $0.9-1.1$. The area of the fitted gaussian along with the statistical uncertainty on the expected signal and background was used to calculate a 90\% C.L.~upper limit on the signal. In the case of \Nuc{Hf}{180} for which the gamma energy of interest is covered by \Nuc{Th}{231} line at 82.1~keV, $S$ was calculated from the difference between the observed and expected backgrounds counts. A summary of the results is shown in Table~\ref{tab:hfLimits}.

\begin{table}[h!]
\centering
\caption{Hf $\alpha$ decays to first excited states of Yb daughter nuclides investigated in this analysis. Gamma energies (E$_\gamma$) of, and yield ($y$) to, the first excited state used to set the limit for each specific decay, experimental lower limits from the analysis presented herein, previous limits~\cite{danevich2020first}, and the theoretical evaluation of half-lives as stated in~\cite{danevich2020first}. Limits are given at 90\% C.L.}
\begin{tabular}{ r | r | r | r | r | r }
Decay& E$_\gamma$~[keV]	& \multicolumn{2}{c|}{Experimental $T_{1/2}$ [a]} & \multicolumn{2}{c}{Theoretical $T_{1/2}$ [a]} \\
Isotope	& \multicolumn{1}{|c|}{($y$ [\%])}	& This work & \multicolumn{1}{|c|}{\cite{danevich2020first}}		& \multicolumn{1}{|c|}{\cite{poenaru1983estimation}}	& \multicolumn{1}{|c}{\cite{buck1991ground,buck1992favoured}}\\ \hline
\Nuc{Hf}{174}	& 84.3 (13.7) 	& $\geq$\scnota{2.8}{16}	& $\geq$\scnota{3.3}{15}	& \scnota{3.0}{18} & \scnota{1.3}{18}\\
\Nuc{Hf}{176}	& 78.7 (10.6) 	& $\geq$\scnota{2.7}{17}	& $\geq$\scnota{3.0}{17}	& \scnota{3.5}{22}	& \scnota{1.3}{22}\\
\Nuc{Hf}{177}	& 78.6 (12.5)	& $\geq$\scnota{1.1}{18} 	& $\geq$\scnota{1.3}{18}	& \scnota{1.2}{24}	& \scnota{9.1}{21}\\
\Nuc{Hf}{178}	& 76.5 (9.6) 	& $\geq$\scnota{1.3}{18}	& $\geq$\scnota{2.0}{17}	& \scnota{8.1}{25}	& \scnota{2.4}{25}\\
\Nuc{Hf}{179}	& 104.5 (26.7)	& $\geq$\scnota{2.7}{18} 	& $\geq$\scnota{2.2}{18}	& \scnota{2.5}{35}& \scnota{2.0}{32}\\
\Nuc{Hf}{180}	& 82.1 (12.4) 	& $\geq$\scnota{4.6}{17}	& $\geq$\scnota{1.0}{18}	& \scnota{4.1}{50}	& \scnota{4.0}{49}\\
\end{tabular}
\label{tab:hfLimits}
\end{table}

No evidence of any alpha decay of Hf isotopes to the first excited state was observed. The half-life results presented here improve the \Nuc{Hf}{174} limit from~\cite{danevich2020first} by a factor of 8.5, which can attributed to the increased detection efficiency and counting time. A search for the decay of \Nuc{Hf}{177} based on the decay of the daughter \Nuc{Yb}{175} ($T_{1/2} = 4.185$~d) at 396~keV yields a less stringent limit at \scnota{3.6}{17}~a than from \Nuc{Hf}{177} directly. The half-life limit of \Nuc{Hf}{180} suffers from the large background of \Nuc{Th}{231} present in the sample. 

Measurement of the decay of \Nuc{Hf}{174} to the first excited state of Yb with the modified HPGe detector is within reach of the theoretically-predicted rate with a factor of 10--100 reduction in backgrounds, which can be achieved with the purification technique of electron beam melting (EBM) as mentioned in~\cite{danevich2020first, pattavina2018innovative}. Measurement of the alpha decays from other Hf isotopes, however, remains unattainable at present with ULB HPGe detectors.

\section{Search for beta decays to excited levels}\label{sec:beta}
In addition to $\alpha$ decay, \Nuc{Hf}{174} is beta unstable with a double beta decay end point energy $Q_{\beta\beta} = 1100.0 \pm 2.3$~keV~\cite{wang2017ame2016}. Two neutrino and neutrinoless $2\varepsilon$ and $\varepsilon\beta^+$ processes in \Nuc{Hf}{174} were investigated using the full 310 day dataset. There are a variety of detectable $\gamma$ and X-rays for these decay transitions. In the process of $2\nu2\varepsilon$ decay,

\begin{equation}\label{eq:2nuecec}
e^- + e^- + (A,Z) \rightarrow (A,Z-2) + 2\nu + 2X,
\end{equation}
the two neutrinos escape the detector and one or both of the two K or L shell X-rays can be measured. Additionally, if the transition is to the first excited state of \Nuc{Yb}{174}, an additional deexcitation $\gamma$ (76.5~keV) can be measured. 

For $0\nu2\varepsilon$ decay,

\begin{equation}\label{eq:2nuecec}
e^- + e^- + (A,Z) \rightarrow (A,Z-2)^{*} \rightarrow (A,Z-2) + \gamma + 2X,
\end{equation} 
the bremsstrahlung photon $\gamma$ is emitted along with two K or L shell X-rays or Auger electrons. This process can also proceed through the first excited state of \Nuc{Yb}{174} with an additional deexcitation gamma that can be measured. The energy of the bremsstrahlung photon is $E_{brem} = Q_{\beta\beta} - E_{b1} - E_{b2} - (E_{\gamma})$ where $E_{b1,b2}$ is the binding energy for the electron shell of the daughter nuclide in the decay mode and, in the transition through the first excited state, $E_\gamma$ is the deexcitation energy~\cite{barabash2020improved}. The 76.5~keV deexcitation gamma can also be used as a signature of this process, however a better sensitivity is achieved using the higher energy bremsstrahlung gamma. The transitions and gamma energies used in this analysis are shown in Table~\ref{tab:betas}. 

The analysis of $2\varepsilon$ decays used two different methods depending on the presence of known background peaks in each energy region of interest. As described in Section~\ref{sec:alpha}, contributions from the background spectrum are only considered when a prominent gamma line exists in the region of interest. In regions with featureless backgrounds the energy spectrum was fit with a linear + gaussian function, and the area of the fitted gaussian along with the statistical uncertainty on the expected signal and background was used to calculate a 90\% C.L.~upper limit on the signal. For the $2L0\nu$ limit, the expected gamma energy of 1003~keV is covered by the large background from the 1001~keV \Nuc{Pa}{234m} line, and the number of excluded events $\lim S$ was calculated from the linear background under the \Nuc{Pa}{234m} peak. The upper limit on the half-life is then calculated as,

\begin{equation}\label{eq:limitB}
	{\rm{lim}}~T_{1/2} = \ln(2)\cdot\mathcal{N}\cdot \eta \cdot t/\lim S,
\end{equation}
where $\mathcal{N}$ is the number of \Nuc{Hf}{174} nuclei, $\eta$ is the FEP detection efficiency, and $t$ is the collection time. The detection efficiency for each mode is obtained from the photon detection efficiency calculated with GEANT4 and, for modes measured from emitted X-rays is combined with the X-ray emission probability. Only the most abundant K shell X-ray line (52.4~keV at 48\%) was used for the $2\nu2K,2\nu KL$ decays to g.s.~and first excited state limits~\cite{firestone1998table}. We also do not consider the $2\nu2L$ process that could be detected through the L shell X-ray line (7.4~keV at 4.9\%) which is below the 18~keV detector energy threshold. The limits calculated, along with previous experimental limits are shown in Table~\ref{tab:betas}. 

The last $\beta\beta$ decay mode investigated here was $K\beta^{+}/L\beta^{+}$ decay,

\begin{equation}\label{eq:ecbplus}
e^- + (A,Z) \rightarrow (A,Z-2) + e^+ + 2\nu,
\end{equation}
where the emitted positron leads to a detectable annihilation photon at 511~keV. The observed counts at 511~keV are consistent with the expected background from \Nuc{Ac}{228} and \Nuc{Tl}{208} in the \Nuc{Th}{232} decay chain. We calculate $\lim S$ based on a fit to the linear background under the 511~keV peak. 

\begin{table}[h!]
\centering
\caption{\Nuc{Hf}{174} electron capture decay modes investigated, $\gamma/X$-ray energies used to set limits for each specific decay mode, efficiency $\eta$, experimental limits from this analysis (90\% C.L.), and existing limits. $K\beta^+$ and $L\beta^+$ limits includes $0\nu$ and $2\nu$ decay modes to both excited and g.s.~transitions. Limits are given at 90\% C.L.}
\begin{tabular}{ r | r | r | r | r}
& && \multicolumn{2}{r}{Experimental $T_{1/2}$ [a]}\\
Decay Mode & $E_{\gamma,X}$ [keV] &	$\eta$ [\%]	&	This work & \cite{danevich2020firstbeta}\\ \hline
$2\nu2K$ (g.s.)		    & 52.4	    & 0.50  & $\geq$ \scnota{1.4}{16}	& $\geq$ \scnota{7.1}{16} \\
$2\nu KL$ (g.s.) 	    & 52.4	    & 0.50  & $\geq$ \scnota{1.4}{16}    & $\geq$ \scnota{3.2}{16} \\
$2\nu2K$ (1st exc.)	 	& 76.5	    & 3.15   & $\geq$ \scnota{7.9}{16}  & -- ~~~~~~ \\
$2\nu2K$ (1st exc.)		& 52.4	    & 0.50  & $\geq$ \scnota{1.4}{16}    & $\geq$ \scnota{5.9}{16} \\
$2\nu KL$ (1st exc.)    & 76.5	    & 3.15 & $\geq$ \scnota{7.9}{16}    &  -- ~~~~~~ \\
$2\nu KL$ (1st exc.)	& 52.4	    & 0.50    & $\geq$ \scnota{1.4}{16}    & $\geq$ \scnota{3.5}{16} \\
$2\nu2L$ (1st exc.) 	& 76.5	    & 3.15 & $\geq$ \scnota{7.9}{16}    & $\geq$ \scnota{3.9}{16} \\ \hline
$0\nu2K$ (g.s.)		    & 977.4		& 7.59 & $\geq$ \scnota{2.7}{18}     & $\geq$ \scnota{5.8}{17} \\ 
$0\nu2K$ (1st exc.)		& 900.9		& 8.01 & $\geq$ \scnota{2.4}{18}    & $\geq$ \scnota{7.1}{17} \\
$0\nu KL$ (g.s.)	    & 1028.9	& 7.32 & $\geq$ \scnota{4.2}{17}    & $\geq$ \scnota{1.9}{18} \\
$0\nu KL$ (1st exc.)	& 952.4		& 7.72 & $\geq$ \scnota{3.1}{17}    & $\geq$ \scnota{6.2}{17} \\
$0\nu LL$ (g.s.) 	    & 1080.4	& 7.09 & $\geq$ \scnota{3.6}{17}    & $\geq$ \scnota{7.8}{17} \\
$0\nu LL$ (1st exc.)	& 1003.9	& 7.45 & $\geq$ \scnota{9.4}{17}    & $\geq$ \scnota{7.2}{17} \\ \hline
$K\beta^+$ ($0\nu+2\nu$)	& 511	& 11.8 & $\geq$ \scnota{5.6}{16}	& $\geq$ \scnota{1.4}{17}\\
$L\beta^+$ ($0\nu+2\nu$)	& 511	& 11.8 & $\geq$ \scnota{5.6}{16}    & $\geq$ \scnota{1.4}{17} \\
\end{tabular}
\label{tab:betas}
\end{table}

No evidence of $\beta\beta$ decays is observed for any mode searched in this study. Due to the higher background rate at 52.4~keV, sensitivity to $2\nu2\varepsilon$ transitions to g.s.~is lower than transitions to the first excited state at 76.5~keV. Limits established in this analysis are comparable to limits presented in \cite{danevich2020firstbeta}. However, due to higher internal backgrounds of the hafnium sample, some limits are less stringent. The inconsistency in improved limits may be explained in the different backgrounds present between this sample and the one used in~\cite{danevich2020firstbeta}.

\section{Search for dark matter-induced deexciations}\label{sec:dm}
The collision of weak-scale dark matter particles with nuclei can give rise to a detectable energy transfer if the cross sections --- and therefore interaction rates --- are high enough. A recent search for collisional deexcitation of metastable nuclear isomer \Nuc{Ta}{180m} induced by Milky Way DM has led to novel constraints on several models of strongly interacting and inelastic DM~\cite{lehnert2019search}. In the latter scenario the dark sector consists of two nearly degenerate states $\chi_1$ and $\chi_2$ with mass splitting $\delta \ll M_\chi$ arising from the introduction of the Majorana mass term~\cite{TuckerSmith:2001hy,Batell:2009vb,Bramante:2020zos}. The lighter $\chi_1$ state usually dominates the DM relic density, and interacts inelastically with Standard Model (SM) particles at tree level: $\chi_1 q \rightarrow \chi_2 q$. In contrast to \Nuc{Ta}{180m}, whose abundance is well measured in natural Ta metal, no natural abundance has been reported for Hf metastable isomers due to their relatively short lifetimes. Therefore, collisional deexcitation cannot be utilized in Hf for dark matter search. However, the excitation energies of the Hf lowest excited states are around 100 keV, and excitation of Hf to these states are still kinematically allowed in collisions with weak-scale dark matter particles. A summary of the excitation energies with spin and parity of the first excitation states are listed in Table~\ref{tab:excitations}. Excited states will promptly fall back to the corresponding ground states via gamma emission. By measuring the activity of the transition lines, we thus obtain a conservative upper bound on the total inelastic scattering rate from DM interactions, and therefore on the DM-nucleon scattering cross section. The measured activities at these gamma ray energies are also shown in Table~\ref{tab:excitations}. In comparison with traditional direct detection experiments, this method appears rather insensitive for two reasons: 1) In collisional excitation dark matter only interacts with the valence nucleons where the coherence enhancement factor is absent, 2) nuclear excitation imposes additional kinematical suppression to overcome the threshold of about $100$ keV. However, we note two advantages: 1) The high mass of Hf means that it is kinematically well-matched to heavier DM candidates, and 2) the region of sensitivity is at higher energies than typically probed in dedicated DM direct detection experiments. We will focus on inelastic dark matter (IDM) candidates which require a very large momentum transfer, outside the analysis region of most direct detection experiments. 

The dark matter-induced collisional excitation rate is given by~\cite{pospelov2019metastable}
\begin{equation}
    R=N_T\dfrac{\rho_\chi}{M_\chi}\int d^3v f(v) \int_{q^2_{\min}}^{q^2_{max}} dq^2 \dfrac{d\sigma_N}{dq^2}S(\vec{q})
\end{equation}
for spin-independent interactions, where $N_T$ is the number of target Hf nuclei of a given isotope in the sample, $\rho_\chi=0.3$ GeV/cm$^3$ is the local dark matter density, and $M_\chi$ is the dark matter mass. We assume a Maxwellian dark matter velocity distribution with $v_0=220$ km/s truncated at the escape velocity $v_{esc}=600$ km/s in the Earth's frame, assuming the Earth's velocity is $v_e=240$ km/s~\cite{monari2018escape}. The minimum/maximum momentum transfer is limited by the velocity of dark matter as
\begin{equation}
    q_{\min/\max}=\mu_{\chi N}v\left[1\mp\sqrt{1-2\dfrac{E_\gamma+\delta }{\mu_{\chi N}v^2}}\right]\,,
\end{equation}
where $\mu_{\chi N}$ is the reduced mass of dark matter and the isotope nucleus, $E_\gamma$ is the excitation energy and $\delta$ is the dark matter mass splitting in case the dark matter particle is scattered into a higher mass state $\chi_2$. The maximum mass splitting that is kinematically accessible is thus 
\begin{equation}
\delta_{\max}=\dfrac{1}{2}\mu_{\chi N}(v_e+v_{esc})^2-E_\gamma\,.    
\end{equation}

\begin{table}[h!]
\centering
\caption{Excitation energies with spin, parity and $B(E2)$ of the first excited states (1st exc.) for transitions from the first excited states to ground states (g.s.) for different Hf isotopes. Data taken from NuDat 2.8~\cite{NuDat2}. The measured gamma ray background near a specific transition energy is given in units of mBq/kg at 68\% C.L. limit in the last column.}

\begin{tabular}{ r | r | r | r | r | r | r}
&\multicolumn{1}{r|}{g.s.}  & \multicolumn{1}{r|}{1st~exc.}& \multicolumn{3}{r|}{Transition} & Bkg.\\ 
Isotope	 &$J^p$  & $J^p$ & $E_\gamma$[keV] & Type & B(E2)[{\it W.u.}]& [mBq/kg]
\\ \hline
\Nuc{Hf}{174}&$0^+$&$2^+$&90.985&E2&152(8)~\cite{Browne:1999xcz}&3.8\\ \hline
\Nuc{Hf}{176}&$0^+$&$2^+$&88.349&E2&183(7)~\cite{Basunia:2006zql} &3.1\\ \hline
\Nuc{Hf}{177}&$7/2^-$&$9/2^-$&112.9500&M1+E2&282(8)~\cite{Kondev:2003whm}&0.9\\ \hline
\Nuc{Hf}{178}&$0^+$&$2^+$&93.1803&E2&160(3)~\cite{Achterberg:2009bix}&2.2\\ \hline
\Nuc{Hf}{179}&$9/2^+$&$11/2^+$&122.7904&M1+E2&245(14)~\cite{Baglin:2009mez}&0.9\\ \hline
\Nuc{Hf}{180}&$0^+$&$2^+$&93.3240&E2&154.8(21)~\cite{McCutchan:2015fnz}&2.2
\end{tabular}
\label{tab:excitations}
\end{table}

In contrast with the coherent elastic scattering case, dark matter that excites the nucleus typically interacts with the few valence nucleons on the surface shell of the nucleus. We include the nuclear transition matrix in the momentum transfer-dependent response function $S(\vec{q})$. As shown in Ref.~\cite{Engel:1999kv}, the nuclear response function can be written as
\begin{equation}
    S(\vec{q})=\sum\limits_L|\langle J_f || j_L(qr)Y_{LM}(\hat{r})||J_i\rangle|^2\,,
\end{equation}
where $J_{i,f}$ are the angular momentum of the ground and excited states, $Y_{LM}$ represents the angular wave function and $j_L$ is the spherical Bessel function. For an $E2$ transition this can be approximated as:
\begin{equation}
    S_{E2}(\vec{q}) = \dfrac{A^2}{Z^2}(2J_i+1)j_2(qR)^2\dfrac{B(E2)}{e^2R^4}\,,
\end{equation}
where the reduced transition probabilities $B(E2)$ are listed in Table~\ref{tab:excitations} in terms of the {\it Weisskopf unit} (W.u.) defined as~\cite{Suhonen:2007zza}
\begin{equation}
    B_{W.u.}(E\lambda)=\dfrac{1.2^{2\lambda}}{4\pi}\left(\dfrac{3}{\lambda+3}\right)^2A^{2\lambda/3}e^2\mathrm{fm}^{2\lambda}\,,
\end{equation}
for a general $E\lambda$ transition. We consider the four-fermion effective interaction with the Lagrangian $L\sim \frac{1}{\Lambda^2}\bar{\chi}_2\chi_1\bar{N}^*N$.  The differential DM-nucleon cross section is given by
\begin{equation}
    \dfrac{d\sigma_n}{dq^2}=\dfrac{\sigma_n}{4v^2\mu_{\chi n}^2}\,.
\end{equation}
We establish our bounds by requiring the excitation rate not to exceed the measured activity near a transition line at 90\% C.L. The measured rate at each energy is determined from the difference in the activity of the sample and background spectra. For energies covered by background peaks, the rate is estimated from the linear fit under the background peaks inferred from the U/Th decay chain (see Sec.~\ref{sec:setup} and~\ref{sec:alpha} for details). This method enables us to explore the deexcitation peaks of \Nuc{Hf}{178} and \Nuc{Hf}{180} at 93~keV, which overlap with the gamma line of \Nuc{Th}{234}.

\begin{figure}[htb!]
\centering
	\includegraphics[width=0.9\textwidth]{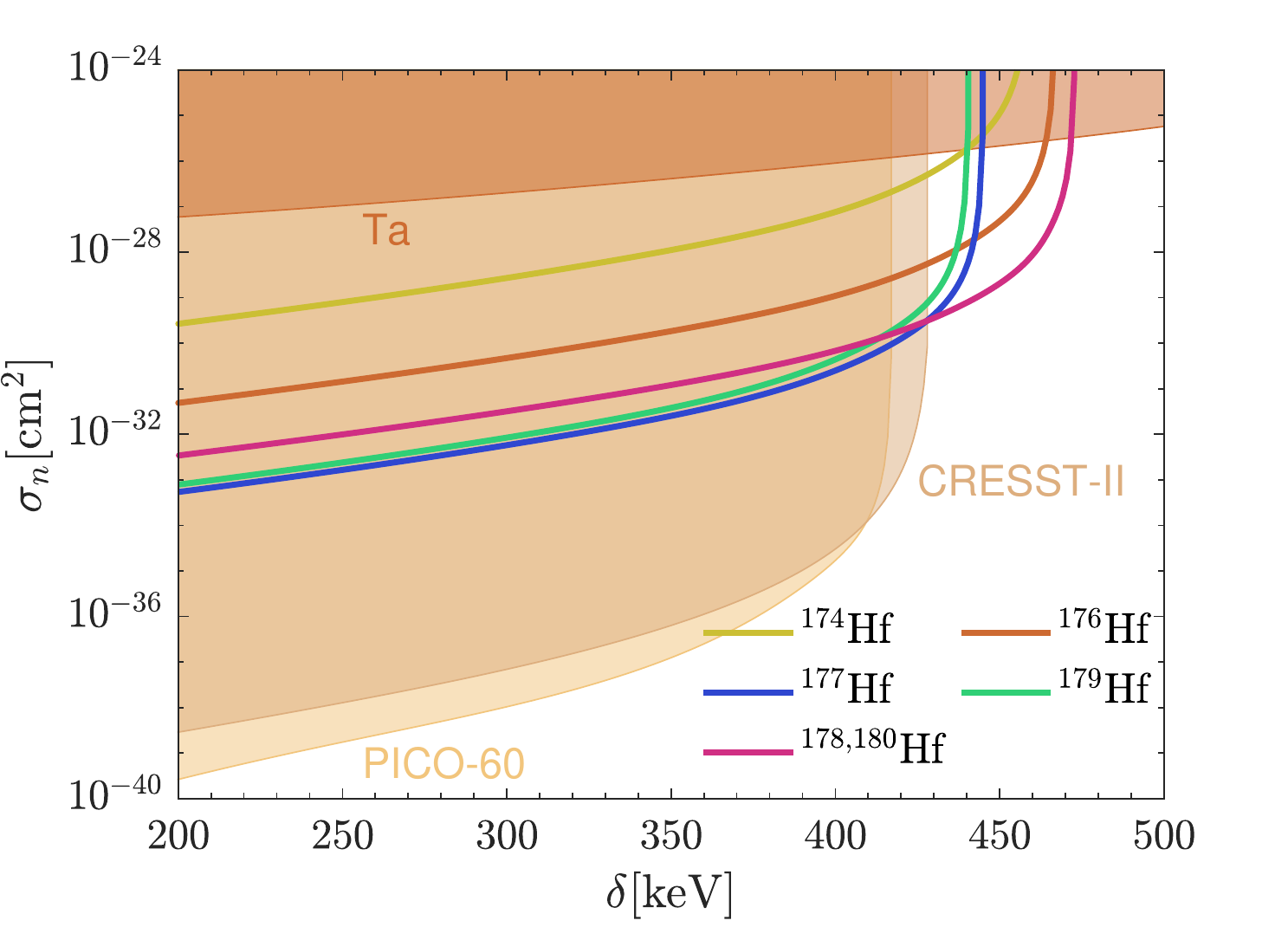}
	\caption{90\% C.L.~bound on inelastic dark matter-nucleon scattering cross section from Hf decay measurements for the dark matter mass $M_\chi=1$~TeV. Limits from PICO-60, CRESST-II~\cite{Bramante:2016rdh} and \Nuc{Ta}{180m} lifetime~\cite{lehnert2019search} have been reproduced by assuming the Earth velocity $v_e=240$~km/s and the escape velocity $v_{esc}=600$~km/s and are shown in shaded regions. The current limits from \Nuc{Hf}{174}, \Nuc{Hf}{176}, \Nuc{Hf}{177}, \Nuc{Hf}{179}, \Nuc{Hf}{178} and \Nuc{Hf}{180} are depicted by solid lines.}
	\label{fig:dmlimit_all}
\end{figure}

We show the limits on IDM-nucleon scattering cross section in Figure~\ref{fig:dmlimit_all} for $M_\chi = 1$~TeV. The benchmark DM mass is chosen following~\cite{Bramante:2016rdh} where $M_\chi$ is much larger than the nucleus mass so that the cross section bound scales simply as $M_\chi$ at high mass range. Existing bounds from PICO~\cite{Amole:2015pla}, CRESST-II~\cite{Angloher:2015ewa} and \Nuc{Ta}{180m} lifetime~\cite{lehnert2019search} are also reproduced with the DM velocity distribution as assumed. Since xenon experiments including XENON1T, LUX and Panda-X only report the recoil energy up to tens of keV, they are only sensitive to the mass splitting $\delta$ below 200~keV. Compared with previous results Hf measurements set the leading bound in the mass splitting between 428~keV and 473~keV. The bound is predominantly set by \Nuc{Hf}{178} and \Nuc{Hf}{180} which share the same transition gamma lines at 93~keV. Below 428~keV mass splitting the Hf bound is dominated by \Nuc{Hf}{177} due to its low background activity, albeit inferior to the PICO and CRESST bounds. 

\begin{figure}[htb!]
\centering
	\includegraphics[width=0.9\textwidth]{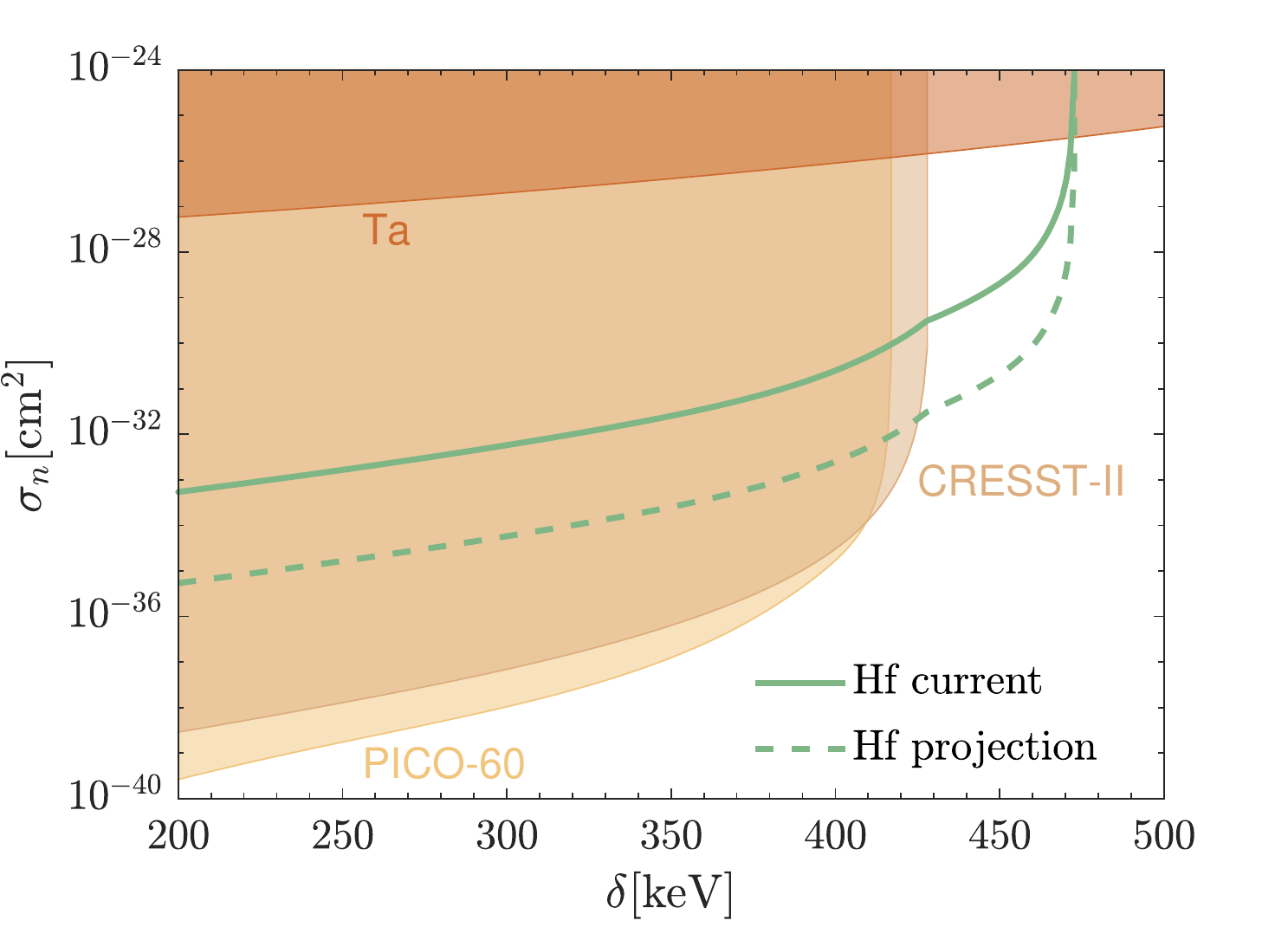}
	\caption{90\% C.L.~bound on inelastic dark matter-nucleon scattering cross section from Hf isotopes. We assume the same dark matter mass and velocity distribution as Figure~\ref{fig:dmlimit_all}. The solid green line shows the current limit combining all Hf isotope bounds and the projection is made in dashed green assuming future purification can reduce the Hf sample background by two orders of magnitude.}
	\label{fig:dmlimit_proj}
\end{figure}

A major limiting factor in the sensitivity in $\sigma_n$ is the high contaminant background rate. The dotted line labeled ``Hf projection'' in Figure~\ref{fig:dmlimit_proj} shows how our bounds could be improved with a reduction of background rates by two orders of magnitude. The projected Hf bound is more stringent than CRESST at $\delta>425$~keV.

We note that even though the bound on the dark matter-nucleon cross section we place is high compared to the ``traditional'' WIMP search region, IDM with such mass splittings are unlikely to be stopped or thermalized due to overburden. This is because the nuclei in the Earth's crust are light enough so that scattering into the heavier dark matter state is not kinematically allowed.

\section{Discussion}\label{sec:disc}
Results presented here improve some alpha and beta decay limits set by~\cite{danevich2020first, danevich2020firstbeta} as a result of the improved detection efficiency for the expected low energy $\gamma/X$-rays. These results are placed using a stock Hf foil with a chemical purity of 99.5\% (excluding Zr) and a concentration of Zr at 1.85\%. As a market-available material without additional purification steps taken, the observed high activity of U/Th decay chains is expected. This corroborates how critically the overall sensitivity of an experimental search depends on optimizing detector setup and geometry.

It should be noted, that secular equilibrium is broken at the level of \Nuc{Ra}{226} in the \Nuc{U}{238} chain, demonstrating that some of the Hf metal production stages and/or preliminary purification stages are strongly selective and take out some nuclides from natural U/Th chains. As shown in~\cite{laubenstein2019new, danevich2020first, bobrov2014purification}, the EBM purification method is very effective in reducing the activity of \Nuc{K}{40} and nuclides in the U/Th chain.

Enrichment of \Nuc{Hf}{174} or \Nuc{Hf}{176} isotopes is not feasible due to the extremely high price of these isotopes ($\sim \$1000/$mg) and limited quantity ($\sim$ g). Therefore, to investigate rare nuclear processes in natural isotopes of Hf to excited states of the daughter nuclides, the combination of ULB HPGe detectors with EBM-purified Hf samples in an optimized geometry is the only path forward. Simply increasing sample mass becomes inefficient as the transmittance of e.g.~84~keV $\gamma$'s in Hf falls below $\mathcal{O}(10^{-9})$ above a 0.2~mm thickness. However, if using the ``source = detector" approach, Cs$_2$HfCl$_6$ crystals with 26\% natural Hf content and excellent scintillation properties are an alternative method. Recently, the direct alpha decay of \Nuc{Hf}{174} was observed with $T_{1/2} = $ \scnota{7}{16}~a using only 7~g of Cs$_2$HfCl$_6$~\cite{caracciolo2020search}. 1-inch-diameter crystals have been grown~\cite{hawrami2020advanced} which can push experimental sensitivity of alpha decays to $T_{1/2}\sim10^{21}$~a with a year of counting and the current level of internal backgrounds~\cite{cardenas2017internal}.

\section{Conclusions}\label{sec:conc}
The first long-term low-background measurement of Hf foil using a custom ULB HPGe detector with optimized Hf sample and detector geometry was performed. Internal backgrounds at the level of Bq/kg in the stock Hf sample were studied in detail and found to be dominated by \Nuc{Th}{234}, \Nuc{Pa}{234m}. Due to its long half-life and exposure on surface, cosmogenically-activated \Nuc{Hf}{181} was additionally present. Further experimental searches with ULB HPGe detectors should make use of the method of increased detection efficiency presented here along with EBM sample purification. 

Experimental limits on $\alpha$ decay of natural Hf isotopes to the first excited state of the daughter Yb nuclides have been placed in the $10^{16}$--$10^{18}$~year range (90\% C.L.), specifically improving the limit of \Nuc{Hf}{174} over the existing limit~\cite{danevich2020first}. Limits on the $0\nu/2\nu~2\varepsilon$ and $\varepsilon \beta^{+}$ decays of \Nuc{Hf}{174} in the range of $10^{16}$--$10^{18}$a (90\% C.L.) are comparable to existing limits~\cite{danevich2020firstbeta}. Finally, the exclusion bounds of heavy inelastic dark matter are improved. For a DM mass of $\sim 1$ TeV, we extend limits on the allowed mass splitting by a modest 10\%. Nonetheless, this serves as an important proof of principle: very small quantities of heavy nuclear isotopes can compete with large-mass purpose-built experiments as probes of cosmological dark matter.

\section*{Acknowledgments}
We thank Joseph Bramante for the helpful discussions. We also acknowledge Harikrishnan Ramani for correspondence on Ta constraints. BB is supported by the Natural Sciences and Engineering Research Council of Canada. NS and ACV are supported by the Arthur B.~McDonald Canadian Astroparticle Physics Research Institute, with equipment funded by the Canada Foundation for Innovation and the Province of Ontario, and housed at the Queen's Centre for Advanced Computing. Research at Perimeter Institute is supported by the Government of Canada through the Department of Innovation, Science, and Economic Development, and by the Province of Ontario.

\bibliographystyle{elsarticle-num-names}
\bibliography{bib.bib}

\begin{thebibliography}{44}
\expandafter\ifx\csname natexlab\endcsname\relax\def\natexlab#1{#1}\fi
\providecommand{\url}[1]{\texttt{#1}}
\providecommand{\href}[2]{#2}
\providecommand{\path}[1]{#1}
\providecommand{\DOIprefix}{doi:}
\providecommand{\ArXivprefix}{arXiv:}
\providecommand{\URLprefix}{URL: }
\providecommand{\Pubmedprefix}{pmid:}
\providecommand{\doi}[1]{\href{http://dx.doi.org/#1}{\path{#1}}}
\providecommand{\Pubmed}[1]{\href{pmid:#1}{\path{#1}}}
\providecommand{\bibinfo}[2]{#2}
\ifx\xfnm\relax \def\xfnm[#1]{\unskip,\space#1}\fi
\bibitem[{Cayrel et~al.(2001)Cayrel, Hill, Beers, Barbuy, Spite, Spite, Plez,
  Andersen, Bonifacio, Francois et~al.}]{cayrel2001measurement}
\bibinfo{author}{R.~Cayrel}, \bibinfo{author}{V.~Hill},
  \bibinfo{author}{T.~Beers}, \bibinfo{author}{B.~Barbuy},
  \bibinfo{author}{M.~Spite}, \bibinfo{author}{F.~Spite},
  \bibinfo{author}{B.~Plez}, \bibinfo{author}{J.~Andersen},
  \bibinfo{author}{P.~Bonifacio}, \bibinfo{author}{P.~Francois}, et~al.,
\newblock \bibinfo{title}{Measurement of stellar age from uranium decay},
\newblock \bibinfo{journal}{Nature} \bibinfo{volume}{409}
  (\bibinfo{year}{2001}) \bibinfo{pages}{691--692}.
\bibitem[{Vergados et~al.(2012)Vergados, Ejiri, and
  {\v{S}}imkovic}]{vergados2012theory}
\bibinfo{author}{J.~Vergados}, \bibinfo{author}{H.~Ejiri},
  \bibinfo{author}{F.~{\v{S}}imkovic},
\newblock \bibinfo{title}{Theory of neutrinoless double-beta decay},
\newblock \bibinfo{journal}{Reports on Progress in Physics}
  \bibinfo{volume}{75} (\bibinfo{year}{2012}) \bibinfo{pages}{106301}.
\bibitem[{Giuliani and Poves(2012)}]{giuliani2012neutrinoless}
\bibinfo{author}{A.~Giuliani}, \bibinfo{author}{A.~Poves},
\newblock \bibinfo{title}{Neutrinoless double-beta decay},
\newblock \bibinfo{journal}{Advances in High Energy Physics}
  \bibinfo{volume}{2012} (\bibinfo{year}{2012}).
\bibitem[{Gomez-Cadenas et~al.(2011)Gomez-Cadenas, Martin-Albo, Mezzetto,
  Monrabal, and Sorel}]{gomez2012search}
\bibinfo{author}{J.~Gomez-Cadenas}, \bibinfo{author}{J.~Martin-Albo},
  \bibinfo{author}{M.~Mezzetto}, \bibinfo{author}{F.~Monrabal},
  \bibinfo{author}{M.~Sorel},
\newblock \bibinfo{title}{The search for neutrinoless double beta decay},
\newblock \bibinfo{journal}{Riv. Nuovo Cim. 35 (2012) 29, arXiv:1109.5515}
  (\bibinfo{year}{2011}).
\bibitem[{Casali et~al.(2016)Casali, Dubovik, Nagorny, Nisi, Orio, Pattavina,
  Pirro, Sch{\"a}ffner, Tupitsyna, and Yakubovskaya}]{casali2016cryogenic}
\bibinfo{author}{N.~Casali}, \bibinfo{author}{A.~Dubovik},
  \bibinfo{author}{S.~Nagorny}, \bibinfo{author}{S.~Nisi},
  \bibinfo{author}{F.~Orio}, \bibinfo{author}{L.~Pattavina},
  \bibinfo{author}{S.~Pirro}, \bibinfo{author}{K.~Sch{\"a}ffner},
  \bibinfo{author}{I.~Tupitsyna}, \bibinfo{author}{A.~Yakubovskaya},
\newblock \bibinfo{title}{Cryogenic detectors for rare alpha decay search: a
  new approach},
\newblock \bibinfo{journal}{Journal of Low Temperature Physics}
  \bibinfo{volume}{184} (\bibinfo{year}{2016}) \bibinfo{pages}{952--957}.
\bibitem[{Casali et~al.(2014)Casali, Nagorny, Orio, Pattavina, Beeman, Bellini,
  Cardani, Dafinei, Di~Domizio, Di~Vacri et~al.}]{casali2014discovery}
\bibinfo{author}{N.~Casali}, \bibinfo{author}{S.~Nagorny},
  \bibinfo{author}{F.~Orio}, \bibinfo{author}{L.~Pattavina},
  \bibinfo{author}{J.~Beeman}, \bibinfo{author}{F.~Bellini},
  \bibinfo{author}{L.~Cardani}, \bibinfo{author}{I.~Dafinei},
  \bibinfo{author}{S.~Di~Domizio}, \bibinfo{author}{M.~Di~Vacri}, et~al.,
\newblock \bibinfo{title}{Discovery of the {151Eu} $\alpha$ decay},
\newblock \bibinfo{journal}{Journal of Physics G: Nuclear and Particle Physics}
  \bibinfo{volume}{41} (\bibinfo{year}{2014}) \bibinfo{pages}{075101}.
\bibitem[{Cozzini et~al.(2004)Cozzini, Angloher, Bucci, von Feilitzsch, Hauff,
  Henry, Jagemann, Jochum, Kraus, Majorovits et~al.}]{cozzini2004detection}
\bibinfo{author}{C.~Cozzini}, \bibinfo{author}{G.~Angloher},
  \bibinfo{author}{C.~Bucci}, \bibinfo{author}{F.~von Feilitzsch},
  \bibinfo{author}{D.~Hauff}, \bibinfo{author}{S.~Henry},
  \bibinfo{author}{T.~Jagemann}, \bibinfo{author}{J.~Jochum},
  \bibinfo{author}{H.~Kraus}, \bibinfo{author}{B.~Majorovits}, et~al.,
\newblock \bibinfo{title}{Detection of the natural $\alpha$ decay of tungsten},
\newblock \bibinfo{journal}{Physical Review C} \bibinfo{volume}{70}
  (\bibinfo{year}{2004}) \bibinfo{pages}{064606}.
\bibitem[{De~Marcillac et~al.(2003)De~Marcillac, Coron, Dambier, Leblanc, and
  Moalic}]{de2003experimental}
\bibinfo{author}{P.~De~Marcillac}, \bibinfo{author}{N.~Coron},
  \bibinfo{author}{G.~Dambier}, \bibinfo{author}{J.~Leblanc},
  \bibinfo{author}{J.-P. Moalic},
\newblock \bibinfo{title}{Experimental detection of $\alpha$-particles from the
  radioactive decay of natural bismuth},
\newblock \bibinfo{journal}{Nature} \bibinfo{volume}{422}
  (\bibinfo{year}{2003}) \bibinfo{pages}{876--878}.
\bibitem[{Laubenstein et~al.(2019)Laubenstein, Lehnert, Nagorny, Nisi, and
  Zuber}]{laubenstein2019new}
\bibinfo{author}{M.~Laubenstein}, \bibinfo{author}{B.~Lehnert},
  \bibinfo{author}{S.~Nagorny}, \bibinfo{author}{S.~Nisi},
  \bibinfo{author}{K.~Zuber},
\newblock \bibinfo{title}{New investigation of half-lives for the decay modes
  of {50V}},
\newblock \bibinfo{journal}{Physical Review C} \bibinfo{volume}{99}
  (\bibinfo{year}{2019}) \bibinfo{pages}{045501}.
\bibitem[{Danevich et~al.(2020)Danevich, Hult, Kasperovych, Kovtun, Kovtun,
  Lutter, Marissens, Polischuk, Stetsenko, and Tretyak}]{danevich2020first}
\bibinfo{author}{F.~Danevich}, \bibinfo{author}{M.~Hult},
  \bibinfo{author}{D.~Kasperovych}, \bibinfo{author}{G.~Kovtun},
  \bibinfo{author}{K.~Kovtun}, \bibinfo{author}{G.~Lutter},
  \bibinfo{author}{G.~Marissens}, \bibinfo{author}{O.~Polischuk},
  \bibinfo{author}{S.~Stetsenko}, \bibinfo{author}{V.~Tretyak},
\newblock \bibinfo{title}{First search for $\alpha$ decays of naturally
  occurring {Hf} nuclides with emission of $\gamma$ quanta},
\newblock \bibinfo{journal}{The European Physical Journal A}
  \bibinfo{volume}{56} (\bibinfo{year}{2020}) \bibinfo{pages}{5}.
\bibitem[{Nagorny et~al.(2020)Nagorny, Laubenstein, and Nisi}]{ptpaper}
\bibinfo{author}{S.~Nagorny}, \bibinfo{author}{M.~Laubenstein},
  \bibinfo{author}{S.~Nisi},
\newblock \bibinfo{title}{Proof of principle of a novel approach to detect rare
  decays using an ultra-low-background high purity germanium detector},
\newblock \bibinfo{journal}{Submitted to JINST}  (\bibinfo{year}{2020}).
\bibitem[{Wang et~al.(2017)Wang, Audi, Kondev, Huang, Naimi, and
  Xu}]{wang2017ame2016}
\bibinfo{author}{M.~Wang}, \bibinfo{author}{G.~Audi},
  \bibinfo{author}{F.~Kondev}, \bibinfo{author}{W.~Huang},
  \bibinfo{author}{S.~Naimi}, \bibinfo{author}{X.~Xu},
\newblock \bibinfo{title}{The {AME2016} atomic mass evaluation (ii). tables,
  graphs and references},
\newblock \bibinfo{journal}{Chinese Physics C} \bibinfo{volume}{41}
  (\bibinfo{year}{2017}) \bibinfo{pages}{030003}.
\bibitem[{Danevich et~al.(2020)Danevich, Hult, Kasperovych, Kovtun, Kovtun,
  Lutter, Marissens, Polischuk, Stetsenko, and Tretyak}]{danevich2020firstbeta}
\bibinfo{author}{F.~Danevich}, \bibinfo{author}{M.~Hult},
  \bibinfo{author}{D.~Kasperovych}, \bibinfo{author}{G.~Kovtun},
  \bibinfo{author}{K.~Kovtun}, \bibinfo{author}{G.~Lutter},
  \bibinfo{author}{G.~Marissens}, \bibinfo{author}{O.~Polischuk},
  \bibinfo{author}{S.~Stetsenko}, \bibinfo{author}{V.~Tretyak},
\newblock \bibinfo{title}{First search for 2$\varepsilon$~and
  $\varepsilon\beta^{+}$~of {174Hf}},
\newblock \bibinfo{journal}{Nuclear Physics A}  (\bibinfo{year}{2020})
  \bibinfo{pages}{121703}.
\bibitem[{Meija et~al.(2016)Meija, Coplen, Berglund, Brand, De~Bi{\`e}vre,
  Gr{\"o}ning, Holden, Irrgeher, Loss, Walczyk et~al.}]{meija2016isotopic}
\bibinfo{author}{J.~Meija}, \bibinfo{author}{T.~B. Coplen},
  \bibinfo{author}{M.~Berglund}, \bibinfo{author}{W.~A. Brand},
  \bibinfo{author}{P.~De~Bi{\`e}vre}, \bibinfo{author}{M.~Gr{\"o}ning},
  \bibinfo{author}{N.~E. Holden}, \bibinfo{author}{J.~Irrgeher},
  \bibinfo{author}{R.~D. Loss}, \bibinfo{author}{T.~Walczyk}, et~al.,
\newblock \bibinfo{title}{Isotopic compositions of the elements 2013 (iupac
  technical report)},
\newblock \bibinfo{journal}{Pure and Applied Chemistry} \bibinfo{volume}{88}
  (\bibinfo{year}{2016}) \bibinfo{pages}{293--306}.
\bibitem[{Pospelov et~al.(2019)Pospelov, Rajendran, and
  Ramani}]{pospelov2019metastable}
\bibinfo{author}{M.~Pospelov}, \bibinfo{author}{S.~Rajendran},
  \bibinfo{author}{H.~Ramani},
\newblock \bibinfo{title}{Metastable nuclear isomers as dark matter
  accelerators},
\newblock \bibinfo{journal}{arXiv preprint arXiv:1907.00011}
  (\bibinfo{year}{2019}).
\bibitem[{hfD(2020)}]{hfDataSheet}
\bibinfo{title}{Certificate available from {alfa.com}}  (\bibinfo{year}{2020}).
\bibitem[{Feldman and Cousins(1998)}]{feldman1998unified}
\bibinfo{author}{G.~J. Feldman}, \bibinfo{author}{R.~D. Cousins},
\newblock \bibinfo{title}{Unified approach to the classical statistical
  analysis of small signals},
\newblock \bibinfo{journal}{Physical Review D} \bibinfo{volume}{57}
  (\bibinfo{year}{1998}) \bibinfo{pages}{3873}.
\bibitem[{Poenaru and Ivascu(1983)}]{poenaru1983estimation}
\bibinfo{author}{D.~Poenaru}, \bibinfo{author}{M.~Ivascu},
\newblock \bibinfo{title}{Estimation of the alpha decay half-lives},
\newblock \bibinfo{journal}{Journal de Physique} \bibinfo{volume}{44}
  (\bibinfo{year}{1983}) \bibinfo{pages}{791--796}.
\bibitem[{Buck et~al.(1991)Buck, Merchant, and Perez}]{buck1991ground}
\bibinfo{author}{B.~Buck}, \bibinfo{author}{A.~Merchant},
  \bibinfo{author}{S.~Perez},
\newblock \bibinfo{title}{Ground state to ground state alpha decays of heavy
  even-even nuclei},
\newblock \bibinfo{journal}{Journal of Physics G: Nuclear and Particle Physics}
  \bibinfo{volume}{17} (\bibinfo{year}{1991}) \bibinfo{pages}{1223}.
\bibitem[{Buck et~al.(1992)Buck, Merchant, and Perez}]{buck1992favoured}
\bibinfo{author}{B.~Buck}, \bibinfo{author}{A.~Merchant},
  \bibinfo{author}{S.~Perez},
\newblock \bibinfo{title}{Favoured alpha decays of odd-mass nuclei},
\newblock \bibinfo{journal}{Journal of Physics G: Nuclear and Particle Physics}
  \bibinfo{volume}{18} (\bibinfo{year}{1992}) \bibinfo{pages}{143}.
\bibitem[{Pattavina et~al.(2018)Pattavina, Laubenstein, Nagorny, Nisi,
  Pagnanini, Pirro, Rusconi, and Sch{\"a}ffner}]{pattavina2018innovative}
\bibinfo{author}{L.~Pattavina}, \bibinfo{author}{M.~Laubenstein},
  \bibinfo{author}{S.~Nagorny}, \bibinfo{author}{S.~Nisi},
  \bibinfo{author}{L.~Pagnanini}, \bibinfo{author}{S.~Pirro},
  \bibinfo{author}{C.~Rusconi}, \bibinfo{author}{K.~Sch{\"a}ffner},
\newblock \bibinfo{title}{An innovative technique for the investigation of the
  4-fold forbidden beta-decay of 50{V}},
\newblock \bibinfo{journal}{The European Physical Journal A}
  \bibinfo{volume}{54} (\bibinfo{year}{2018}) \bibinfo{pages}{79}.
\bibitem[{Barabash et~al.(2020)Barabash, Brudanin, Klimenko, Konovalov,
  Rakhimov, Rukhadze, Rukhadze, Shitov, Stekl, Warot
  et~al.}]{barabash2020improved}
\bibinfo{author}{A.~Barabash}, \bibinfo{author}{V.~Brudanin},
  \bibinfo{author}{A.~Klimenko}, \bibinfo{author}{S.~Konovalov},
  \bibinfo{author}{A.~Rakhimov}, \bibinfo{author}{E.~Rukhadze},
  \bibinfo{author}{N.~Rukhadze}, \bibinfo{author}{Y.~A. Shitov},
  \bibinfo{author}{I.~Stekl}, \bibinfo{author}{G.~Warot}, et~al.,
\newblock \bibinfo{title}{Improved limits on $\beta$+ {EC} and {ECEC} processes
  in 74{Se}},
\newblock \bibinfo{journal}{Nuclear Physics A}  (\bibinfo{year}{2020})
  \bibinfo{pages}{121697}.
\bibitem[{Firestone and Shirley(1998)}]{firestone1998table}
\bibinfo{author}{R.~B. Firestone}, \bibinfo{author}{V.~S. Shirley},
\newblock \bibinfo{title}{Table of isotopes, 2 volume set},
\newblock \bibinfo{journal}{Table of Isotopes, 2 Volume Set, by Richard B.
  Firestone, Virginia S. Shirley (Editor), pp. 3168. ISBN 0-471-33056-6.
  Wiley-VCH, December 1998.}  (\bibinfo{year}{1998}) \bibinfo{pages}{3168}.
\bibitem[{Lehnert et~al.(2019)Lehnert, Ramani, Hult, Lutter, Pospelov,
  Rajendran, and Zuber}]{lehnert2019search}
\bibinfo{author}{B.~Lehnert}, \bibinfo{author}{H.~Ramani},
  \bibinfo{author}{M.~Hult}, \bibinfo{author}{G.~Lutter},
  \bibinfo{author}{M.~Pospelov}, \bibinfo{author}{S.~Rajendran},
  \bibinfo{author}{K.~Zuber},
\newblock \bibinfo{title}{Search for dark matter induced de-excitation of
  ${}^{180m}${Ta}},
\newblock \bibinfo{journal}{arXiv preprint arXiv:1911.07865}
  (\bibinfo{year}{2019}).
\bibitem[{Tucker-Smith and Weiner(2001)}]{TuckerSmith:2001hy}
\bibinfo{author}{D.~Tucker-Smith}, \bibinfo{author}{N.~Weiner},
\newblock \bibinfo{title}{{Inelastic dark matter}},
\newblock \bibinfo{journal}{Phys. Rev. D} \bibinfo{volume}{64}
  (\bibinfo{year}{2001}) \bibinfo{pages}{043502}.
  \DOIprefix\doi{10.1103/PhysRevD.64.043502}.
  \href{http://arxiv.org/abs/hep-ph/0101138}{{\tt arXiv:hep-ph/0101138}}.
\bibitem[{Batell et~al.(2009)Batell, Pospelov, and Ritz}]{Batell:2009vb}
\bibinfo{author}{B.~Batell}, \bibinfo{author}{M.~Pospelov},
  \bibinfo{author}{A.~Ritz},
\newblock \bibinfo{title}{{Direct Detection of Multi-component Secluded
  WIMPs}},
\newblock \bibinfo{journal}{Phys. Rev. D} \bibinfo{volume}{79}
  (\bibinfo{year}{2009}) \bibinfo{pages}{115019}.
  \DOIprefix\doi{10.1103/PhysRevD.79.115019}.
  \href{http://arxiv.org/abs/0903.3396}{{\tt arXiv:0903.3396}}.
\bibitem[{Bramante and Song(2020)}]{Bramante:2020zos}
\bibinfo{author}{J.~Bramante}, \bibinfo{author}{N.~Song},
\newblock \bibinfo{title}{{Electric But Not Eclectic: Thermal Relic Dark Matter
  for the XENON1T Excess}},
\newblock \bibinfo{journal}{Phys. Rev. Lett.} \bibinfo{volume}{125}
  (\bibinfo{year}{2020}) \bibinfo{pages}{161805}.
  \DOIprefix\doi{10.1103/PhysRevLett.125.161805}.
  \href{http://arxiv.org/abs/2006.14089}{{\tt arXiv:2006.14089}}.
\bibitem[{Monari et~al.(2018)Monari, Famaey, Carrillo, Piffl, Steinmetz, Wyse,
  Anders, Chiappini, and Janssen}]{monari2018escape}
\bibinfo{author}{G.~Monari}, \bibinfo{author}{B.~Famaey},
  \bibinfo{author}{I.~Carrillo}, \bibinfo{author}{T.~Piffl},
  \bibinfo{author}{M.~Steinmetz}, \bibinfo{author}{R.~F. Wyse},
  \bibinfo{author}{F.~Anders}, \bibinfo{author}{C.~Chiappini},
  \bibinfo{author}{K.~Janssen},
\newblock \bibinfo{title}{The escape speed curve of the galaxy obtained from
  {Gaia DR2} implies a heavy milky way},
\newblock \bibinfo{journal}{Astronomy \& Astrophysics} \bibinfo{volume}{616}
  (\bibinfo{year}{2018}) \bibinfo{pages}{L9}.
\bibitem[{NuD(2020)}]{NuDat2}
\bibinfo{title}{{NuDat} 2.8}, \bibinfo{year}{2020}. \URLprefix
  \url{https://www.nndc.bnl.gov/nudat2/}.
\bibitem[{Browne and Junde(1999)}]{Browne:1999xcz}
\bibinfo{author}{E.~Browne}, \bibinfo{author}{H.~Junde},
\newblock \bibinfo{title}{{Nuclear Data Sheets for A = 174}},
\newblock \bibinfo{journal}{Nucl. Data Sheets} \bibinfo{volume}{87}
  (\bibinfo{year}{1999}) \bibinfo{pages}{15--176}.
  \DOIprefix\doi{10.1006/ndsh.1999.0015}.
\bibitem[{Basunia(2006)}]{Basunia:2006zql}
\bibinfo{author}{M.~Basunia},
\newblock \bibinfo{title}{{Nuclear Data Sheets for A = 176}},
\newblock \bibinfo{journal}{Nucl. Data Sheets} \bibinfo{volume}{107}
  (\bibinfo{year}{2006}) \bibinfo{pages}{791--1026}.
  \DOIprefix\doi{10.1016/j.nds.2006.03.001}.
\bibitem[{Kondev(2003)}]{Kondev:2003whm}
\bibinfo{author}{F.~Kondev},
\newblock \bibinfo{title}{{Nuclear Data Sheets for A = 177}},
\newblock \bibinfo{journal}{Nucl. Data Sheets} \bibinfo{volume}{98}
  (\bibinfo{year}{2003}) \bibinfo{pages}{801--1095}.
  \DOIprefix\doi{10.1006/ndsh.2003.0006}.
\bibitem[{Achterberg et~al.(2009)Achterberg, Capurro, and
  Marti}]{Achterberg:2009bix}
\bibinfo{author}{E.~Achterberg}, \bibinfo{author}{O.~Capurro},
  \bibinfo{author}{G.~Marti},
\newblock \bibinfo{title}{{Nuclear Data Sheets for A = 178}},
\newblock \bibinfo{journal}{Nucl. Data Sheets} \bibinfo{volume}{110}
  (\bibinfo{year}{2009}) \bibinfo{pages}{1473--1688}.
  \DOIprefix\doi{10.1016/j.nds.2009.05.002}.
\bibitem[{Baglin(2009)}]{Baglin:2009mez}
\bibinfo{author}{C.~M. Baglin},
\newblock \bibinfo{title}{{Nuclear Data Sheets for A = 179}},
\newblock \bibinfo{journal}{Nucl. Data Sheets} \bibinfo{volume}{110}
  (\bibinfo{year}{2009}) \bibinfo{pages}{265--506}.
  \DOIprefix\doi{10.1016/j.nds.2009.01.001}.
\bibitem[{McCutchan(2015)}]{McCutchan:2015fnz}
\bibinfo{author}{E.~McCutchan},
\newblock \bibinfo{title}{{Nuclear Data Sheets for A = 180}},
\newblock \bibinfo{journal}{Nucl. Data Sheets} \bibinfo{volume}{126}
  (\bibinfo{year}{2015}) \bibinfo{pages}{151--372}.
  \DOIprefix\doi{10.1016/j.nds.2015.05.002}.
\bibitem[{Engel and Vogel(2000)}]{Engel:1999kv}
\bibinfo{author}{J.~Engel}, \bibinfo{author}{P.~Vogel},
\newblock \bibinfo{title}{{Neutralino inelastic scattering with subsequent
  detection of nuclear gamma-rays}},
\newblock \bibinfo{journal}{Phys. Rev.} \bibinfo{volume}{D61}
  (\bibinfo{year}{2000}) \bibinfo{pages}{063503}.
  \DOIprefix\doi{10.1103/PhysRevD.61.063503}.
  \href{http://arxiv.org/abs/hep-ph/9910409}{{\tt arXiv:hep-ph/9910409}}.
\bibitem[{Suhonen(2007)}]{Suhonen:2007zza}
\bibinfo{author}{J.~Suhonen}, \bibinfo{title}{{From Nucleons to Nucleus}},
  Theoretical and Mathematical Physics, \bibinfo{publisher}{Springer},
  \bibinfo{address}{Berlin, Germany}, \bibinfo{year}{2007}.
  \DOIprefix\doi{10.1007/978-3-540-48861-3}.
\bibitem[{Bramante et~al.(2016)Bramante, Fox, Kribs, and
  Martin}]{Bramante:2016rdh}
\bibinfo{author}{J.~Bramante}, \bibinfo{author}{P.~J. Fox},
  \bibinfo{author}{G.~D. Kribs}, \bibinfo{author}{A.~Martin},
\newblock \bibinfo{title}{{Inelastic frontier: Discovering dark matter at high
  recoil energy}},
\newblock \bibinfo{journal}{Phys. Rev.} \bibinfo{volume}{D94}
  (\bibinfo{year}{2016}) \bibinfo{pages}{115026}.
  \DOIprefix\doi{10.1103/PhysRevD.94.115026}.
  \href{http://arxiv.org/abs/1608.02662}{{\tt arXiv:1608.02662}}.
\bibitem[{Amole et~al.(2016)}]{Amole:2015pla}
\bibinfo{author}{C.~Amole}, et~al. (\bibinfo{collaboration}{PICO}),
\newblock \bibinfo{title}{{Dark matter search results from the PICO-60 CF$_3$I
  bubble chamber}},
\newblock \bibinfo{journal}{Phys. Rev.} \bibinfo{volume}{D93}
  (\bibinfo{year}{2016}) \bibinfo{pages}{052014}.
  \DOIprefix\doi{10.1103/PhysRevD.93.052014}.
  \href{http://arxiv.org/abs/1510.07754}{{\tt arXiv:1510.07754}}.
\bibitem[{Angloher et~al.(2016)}]{Angloher:2015ewa}
\bibinfo{author}{G.~Angloher}, et~al. (\bibinfo{collaboration}{CRESST}),
\newblock \bibinfo{title}{{Results on light dark matter particles with a
  low-threshold CRESST-II detector}},
\newblock \bibinfo{journal}{Eur. Phys. J.} \bibinfo{volume}{C76}
  (\bibinfo{year}{2016}) \bibinfo{pages}{25}.
  \DOIprefix\doi{10.1140/epjc/s10052-016-3877-3}.
  \href{http://arxiv.org/abs/1509.01515}{{\tt arXiv:1509.01515}}.
\bibitem[{Bobrov et~al.(2014)Bobrov, Dmitrenko, Koblik, Lavrinenko,
  Laubenstein, Nagorny, Pylypenko, Stadnik, Tantsyura, and
  Virich}]{bobrov2014purification}
\bibinfo{author}{Y.~P. Bobrov}, \bibinfo{author}{A.~Dmitrenko},
  \bibinfo{author}{D.~Koblik}, \bibinfo{author}{S.~Lavrinenko},
  \bibinfo{author}{M.~Laubenstein}, \bibinfo{author}{S.~Nagorny},
  \bibinfo{author}{M.~Pylypenko}, \bibinfo{author}{Y.~S. Stadnik},
  \bibinfo{author}{I.~Tantsyura}, \bibinfo{author}{V.~Virich},
\newblock \bibinfo{title}{Purification of vanadium by electron-beam melting},
\newblock \bibinfo{journal}{Problems of Atomic Science and Technology}
  (\bibinfo{year}{2014}).
\bibitem[{Caracciolo et~al.(2020)Caracciolo, Nagorny, Belli, Bernabei,
  Cappella, Cerulli, Incicchitti, Laubenstein, Merlo, Nisi
  et~al.}]{caracciolo2020search}
\bibinfo{author}{V.~Caracciolo}, \bibinfo{author}{S.~Nagorny},
  \bibinfo{author}{P.~Belli}, \bibinfo{author}{R.~Bernabei},
  \bibinfo{author}{F.~Cappella}, \bibinfo{author}{R.~Cerulli},
  \bibinfo{author}{A.~Incicchitti}, \bibinfo{author}{M.~Laubenstein},
  \bibinfo{author}{V.~Merlo}, \bibinfo{author}{S.~Nisi}, et~al.,
\newblock \bibinfo{title}{Search for $\alpha$ decay of naturally occurring
  {Hf}-nuclides using a {Cs2HfCl6} scintillator},
\newblock \bibinfo{journal}{Nuclear Physics A}  (\bibinfo{year}{2020})
  \bibinfo{pages}{121941}.
\bibitem[{Hawrami et~al.(2020)Hawrami, Ariesanti, Buliga, Matei, Motakef, and
  Burger}]{hawrami2020advanced}
\bibinfo{author}{R.~Hawrami}, \bibinfo{author}{E.~Ariesanti},
  \bibinfo{author}{V.~Buliga}, \bibinfo{author}{L.~Matei},
  \bibinfo{author}{S.~Motakef}, \bibinfo{author}{A.~Burger},
\newblock \bibinfo{title}{Advanced high-performance large diameter {{Cs2HfCl6}
  (CHC)} and mixed halides scintillator},
\newblock \bibinfo{journal}{Journal of Crystal Growth} \bibinfo{volume}{533}
  (\bibinfo{year}{2020}) \bibinfo{pages}{125473}.
\bibitem[{Cardenas et~al.(2017)Cardenas, Burger, DiVacri, Goodwin, Groza,
  Laubenstein, Nagorny, Nisi, and Rowe}]{cardenas2017internal}
\bibinfo{author}{C.~Cardenas}, \bibinfo{author}{A.~Burger},
  \bibinfo{author}{M.~DiVacri}, \bibinfo{author}{B.~Goodwin},
  \bibinfo{author}{M.~Groza}, \bibinfo{author}{M.~Laubenstein},
  \bibinfo{author}{S.~Nagorny}, \bibinfo{author}{S.~Nisi},
  \bibinfo{author}{E.~Rowe},
\newblock \bibinfo{title}{Internal contamination of the {Cs2HfCl6} crystal
  scintillator},
\newblock \bibinfo{journal}{Nuclear Instruments and Methods in Physics Research
  Section A: Accelerators, Spectrometers, Detectors and Associated Equipment}
  \bibinfo{volume}{872} (\bibinfo{year}{2017}) \bibinfo{pages}{23--27}.

\end{thebibliography}

\end{document}